\documentclass[aip, amsmath,amssymb, reprint]{revtex4-1}

\usepackage{graphicx}
\usepackage{dcolumn}
\usepackage{bm}

\usepackage[utf8]{inputenc}
\usepackage[T1]{fontenc}
\usepackage{mathptmx}
\usepackage{etoolbox}

\usepackage{natbib}
\usepackage{lineno}
\usepackage{siunitx}
\usepackage{braket}
\usepackage{ulem}

\def\u#1{_{\rm #1}}

\DeclareRobustCommand{\erase}{\bgroup\markoverwith{\textcolor{red}{\rule[.5ex]{2pt}{0.4pt}}}\ULon}

\makeatletter
\def\@email#1#2{%
 \endgroup
 \patchcmd{\titleblock@produce}
  {\frontmatter@RRAPformat}
  {\frontmatter@RRAPformat{\produce@RRAP{*#1\href{mailto:#2}{#2}}}\frontmatter@RRAPformat}
  {}{}
}%
\makeatother
\begin{document}

\preprint{AIP/123-QED}

\title{$1\times N$ DWDM channel selective quantum frequency conversion}
\author{T. Arizono}
 \affiliation{Graduate School of Engineering Science, Osaka University, Osaka 560-8531, Japan}
\author{T. Kobayashi}%
 \affiliation{Graduate School of Engineering Science, Osaka University, Osaka 560-8531, Japan}
 \affiliation{Center for Quantum Information and Quantum Biology, 
Osaka University, Osaka 560-8531, Japan}
\author{S. Miki}
\affiliation{%
Advanced ICT Research Institute, National Institute of Information and Communications Technology, Hyogo, 651-2492, Japan
}%
\author{H. Terai}
\affiliation{%
Advanced ICT Research Institute, National Institute of Information and Communications Technology, Hyogo, 651-2492, Japan
}%
\author{T. Kodama}
\affiliation{%
Advanced ICT Research Institute, National Institute of Information and Communications Technology, Hyogo, 651-2492, Japan
}%
\affiliation{%
Hamamatsu Photonics K.K., Shizuoka, Japan
}%
\author{H. Shimoi}
\affiliation{%
Hamamatsu Photonics K.K., Shizuoka, Japan
}%
\author{T. Yamamoto}
 \affiliation{Graduate School of Engineering Science, Osaka University, Osaka 560-8531, Japan}
 \affiliation{Center for Quantum Information and Quantum Biology, 
Osaka University, Osaka 560-8531, Japan}
\author{R. Ikuta}
 \email{ikuta.rikizo.es@osaka-u.ac.jp}
 \affiliation{Graduate School of Engineering Science, Osaka University, Osaka 560-8531, Japan}
 \affiliation{Center for Quantum Information and Quantum Biology, 
Osaka University, Osaka 560-8531, Japan}

\begin{abstract}
Dense Wavelength Division Multiplexing~(DWDM) is a key technology for realizing high-capacity and flexible quantum communication networks.
In addition, to realize the emerging quantum internet, quantum frequency conversion is also essential for bridging different quantum systems over optical fiber networks.
In this work, we demonstrate a channel-selective quantum frequency conversion (CS-QFC), which allows active selection of the frequency of the converted photon from multiple DWDM channels.
The \SI{2.5}{THz} bandwidth of our CS-QFC system shows the ability to establish a 100-ch DWDM dynamic link from a single quantum system.
It promises to increase the diversity of the quantum network.
\end{abstract}

\maketitle

\section{\label{sec:level1}introduction}
Frequency multiplexing in optical fiber networks is essential for high-speed and multi-user quantum communications. 
The generation and distribution of frequency-multiplexed entangled photon pairs based on broadband spontaneous parametric downconversion~(SPDC),
comparable to dense wavelength division multiplexing~(DWDM) technologies,
has been widely studied~\cite{Lim2007-op,Lim2008-qf,Lim2008-jn,Zhou2013-zk,Ghalbouni2013-cv,Aktas2016-be,Wengerowsky2018-oz,Joshi2020-jz,Mueller2024}. 
However, SPDC-based photon sources will not always be applicable to the quantum internet~\cite{Kimble2008,Wehner2018} which includes photonic systems and heterogeneous quantum matter systems such as atoms, ions, semiconductors, and superconductors. 
Experimental research to establish an entanglement between distant quantum matter systems via optical fiber-based quantum communication is increasingly active. 
In the studies, an efficient distribution of photons through optical fibers 
was achieved using quantum frequency conversion~(QFC), which converts a wavelength of photons emitted from a quantum matter to a wavelength in the telecom band without destroying the entanglement~\cite{Ikuta2011,Ikuta2018,Bock2018,Tchebotareva2019,Krutyanskiy2023,Bersin2024,Zhou2024,Liu2024,Knaut2024}. 
The process of QFC is based on nonlinear optical interaction 
with the use of a strong pump light at a frequency 
corresponding to the difference between the frequencies of the input and the converted photons. 
The QFC to a single telecom wavelength has been used in each experiment so far. 
However, in order to achieve any-to-any quantum communication between $N$ parties assigned to $N$ frequency channels in the DWDM network, based on the two-photon or single-photon interference at the midpoints, 
it is necessary to actively route each photon to the desired frequency channel 
so that the photons meet at the appropriate midpoints for the Bell measurement. 

For this purpose, we demonstrate a QFC where 
the frequency of the converted photon can be actively selected 
from the DWDM channels. 
We call it channel selective QFC~(CS-QFC) hereafter. 
The conceptual figure of the application of the CS-QFC to the multiparty quantum communication is shown in Fig.~\ref{fig:setup}(a). 
The CS-QFC is achieved by selecting one of the multiple pump lasers, each frequency detuned from the others, for the nonlinear optical interaction. 
In the experiment, the CS-QFC is based on a second-order optical nonlinearity in a periodically poled lithium niobate~(PPLN) waveguide 
used for conventional QFC experiments. 
A single photon at \SI{780}{nm} is input to the CS-QFC. 
Using one of seven pump lights around \SI{1580}{nm}
with a frequency detuning of \SI{25}{GHz}, 
the signal photon is converted to wavelengths of around \SI{1540}{nm}
corresponding to the pump frequencies, and is distributed to 
different output channels of the DWDM demultiplexer~(DeMux) without a serious channel crosstalk. 
The frequency range available for multiplexing demonstrated here is investigated to be over \SI{2.5}{THz}. 
This shows the possibility of 100-mode frequency multiplexing for channel spacing of \SI{25}{GHz}. 

\section{Theory}
The theory of the CS-QFC in this paper follows conventional QFCs based on difference frequency generation using a pump light at a single frequency mode, which we briefly review~\cite{Ikuta2011}. 
The Hamiltonian of QFC from a signal mode at angular frequency $\omega_{s}$ to a converted mode at $\omega_{c}$ using a sufficiently strong pump light at $\omega_{p}=\omega_{s}-\omega_{c}$ is 
$H=i \hbar (\xi^* a_s a_{c}^\dagger - \xi a_s^\dagger a_{c})$, 
where $a_s$ and $a_c$ are annihilation operators of the signal and converted modes. 
$\xi$ is a coupling constant proportional to the complex amplitude of the pump light. 
By solving the Heisenberg equation, 
the time evolution of the converted mode with interaction time $\tau$ is obtained, 
which is regarded as the converted mode coming from the nonlinear optical medium. 
The annihilation operator of the mode denoted by $b_{c}$ is described by 
$b_{c} = e^{-i\phi}\sin(\theta/2) a_{s} +\cos(\theta/2) a_{c}$, 
where $\phi$ is the phase of the pump light and $\theta/2 = |\xi|\tau$. 

For the CS-QFC, 
we consider multiple pump lights at angular frequencies of $\omega_{p,i}$ 
for $1\leq i\leq N$ including multiple communication round 
as shown in Figs.~\ref{fig:setup}(a) and (b). 
When the pump light at $\omega_{p,i}$ is used in the $T$-th round of QFC, 
the annihilation operator $b_{c,T,i}$ of the converted mode 
at angular frequency $\omega_{c,i} = \omega_s - \omega_{p,i}$ 
from the nonlinear optical medium on $T$-th round is 
\begin{align}
b_{c,T,i} = e^{-i\phi_i}\sin(\theta_i/2) a_{s,T,i} +\cos(\theta_i/2) a_{c,T,i}, 
\label{eq:qfc}
\end{align}
where 
$[a_{s,T,i},a_{s,T',j}^\dagger]=[a_{c,T,i},a_{c,T',j}^\dagger]=\delta_{T,T'}\delta_{i,j}$, 
leading to $[b_{c,T,i},b_{c,T',j}^\dagger]=\delta_{T,T'}\delta_{i,j}$. 
$\phi_i$ and $\theta_i$ depend on the complex amplitude of the pump light at $\omega_{p,i}$. 
From Eq.~(\ref{eq:qfc}), 
the deterministic QFC of a photon from $\omega_s$ to $\omega_{c,i}$ 
is achieved for each round by satisfying $\theta_i=\pi$.

The above description of QFC holds 
for the region of frequencies satisfying 
the phase-matching condition 
$\Delta k = k(\omega_s) -k(\omega_{p,i})-k(\omega_{c,i})=0$, 
where $k(\omega)$ is the wave number of photons at angular frequency $\omega$. 
The larger value of $|\Delta k|$ leads to the lower value of the maximum conversion efficiency. 
In other words, the number $N$ of the possible selections for the converted modes 
by the CS-QFC is limited by the phase-matching bandwidth. 
The phase-matching bandwidth related to the QFC is the same as that of 
SPDC photon pairs at $\omega_{p,i}$ and $\omega_{c,i}$ pumped by a laser light at $\omega_{s}$. 
In typical experiments related to SPDC photon pairs at telecom wavelengths 
using PPLN waveguides, broad bandwidths larger than THz have been observed\cite{Zhang2021}. 

For CS-QFC, only one pump light per round is used. 
If more than two pump lights are simultaneously input to the nonlinear optical medium, interactions between multiple signal modes and multiple converted modes with frequency spacings determined by the pump lights are induced. 
While such an interaction may be used for manipulating single-photon frequency combs, it will not be unsuitable for the deterministic frequency conversion of a photon at a single frequency mode to a different single frequency mode. 

\section{experimental setup}
\begin{figure*}[t]
 \begin{center}
      \scalebox{0.9}{\includegraphics{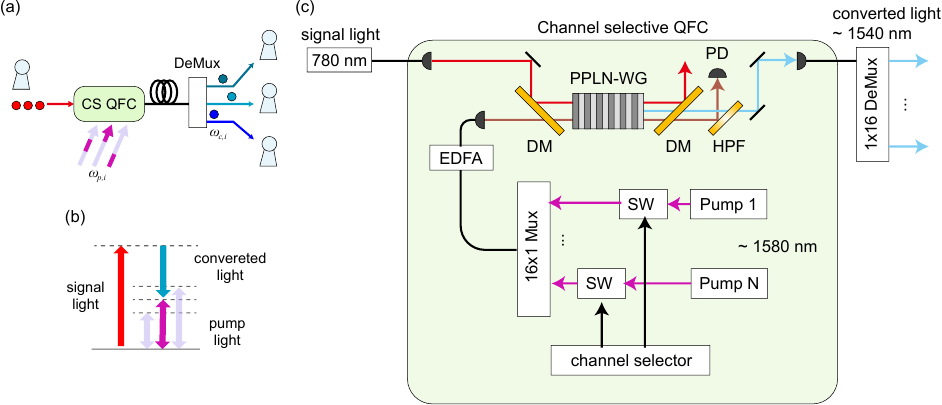}}
    \caption{
      (a) Concept of CS-QFC.
      (b) Energy diagram of CS-QFC.
      (c) Experimental setup.
    }
    \label{fig:setup}
 \end{center}
\end{figure*}
The experimental setup for the CS-QFC is shown in Fig.~\ref{fig:setup}(c). 
The CS-QFC has an input port for a signal light at \SI{780}{nm}. 
The pump light for CS-QFC is prepared by a multi-channel wavelength tunable laser. 
The frequencies of the pump lasers can be independently set from 190.7 to \SI{186.7}{THz} corresponding to the wavelengths from 1572.1 to \SI{1605.7}{nm}. 
Each pump laser is connected to an optical shutter 
based on the InP/InGaAsP multiple quantum well~(Thorlabs; BOA1004PXS) 
driven by a current source controller~(Thorlabs; CLD1015) 
which has a switching speed of about \SI{4}{\mu s}. 
We can turn on and off the output power of each pump laser 
by modulating the voltage to the current source controller.
The output power of the laser from the optical shutter can be adjusted 
by the operating current applied to the shutter. 
Each output light from the optical shutters is connected to 
a 16x1 frequency multiplexer~(Mux) whose channel spacing is \SI{25}{GHz}. 
The multiplexed light from the Mux is amplified 
by an erbium-doped fiber amplifier~(EDFA) to a maximum power of \SI{500}{mW}.
The pump current of the EDFA is driven in the auto current control mode. 
With the proper setting of the current values applied to the optical shutters, 
the pump power after the amplification at each pump frequency is adjusted 
to a value giving a maximum conversion efficiency of CS-QFC. 

The amplified pump light is combined with the signal light at \SI{780}{nm}, 
and they are coupled to the PPLN-WG. The length of PPLN-WG is \SI{40}{mm}. 
The DFG process in the PPLN-WG produces a frequency-converted light around \SI{1540}{nm}. 
The \SI{1540}{nm} light coming from the PPLN-WG is separated from the signal light and the pump light by using a dichroic mirror~(DM), a high-pass filter~(HPF), respectively.
Then, the converted light is coupled to a single-mode fiber~(SMF) 
which is followed by a 1x16 frequency DeMux and a measurement apparatus such as an optical power meter, photo detectors~(PDs), and superconducting nanostrip single photon detectrors~(SNSPDs) developed by Hamamatsu photonics and NICT.

In the demonstration of the CS-QFC, a \SI{780}{nm} photon is prepared as the input signal photon
by the SPDC process using another PPLN-WG~(not shown). 
The pump light at \SI{517}{nm} for SPDC generates \SI{780}{nm} and \SI{1540}{nm} photons. 
The \SI{1540}{nm} photon heralds the \SI{780}{nm} photon by the photon detection with a SNSPD after passing through a bandpass filter with a bandwidth of \SI{10}{GHz}. 
The frequency-converted photon of the heralded \SI{780}{nm} photon is detected by SNSPDs. 

\section{Experimental results}
\subsection{Characterization of channel-selective frequency conversion}
\begin{figure}[t]
  \centering
  \includegraphics[width=7cm]{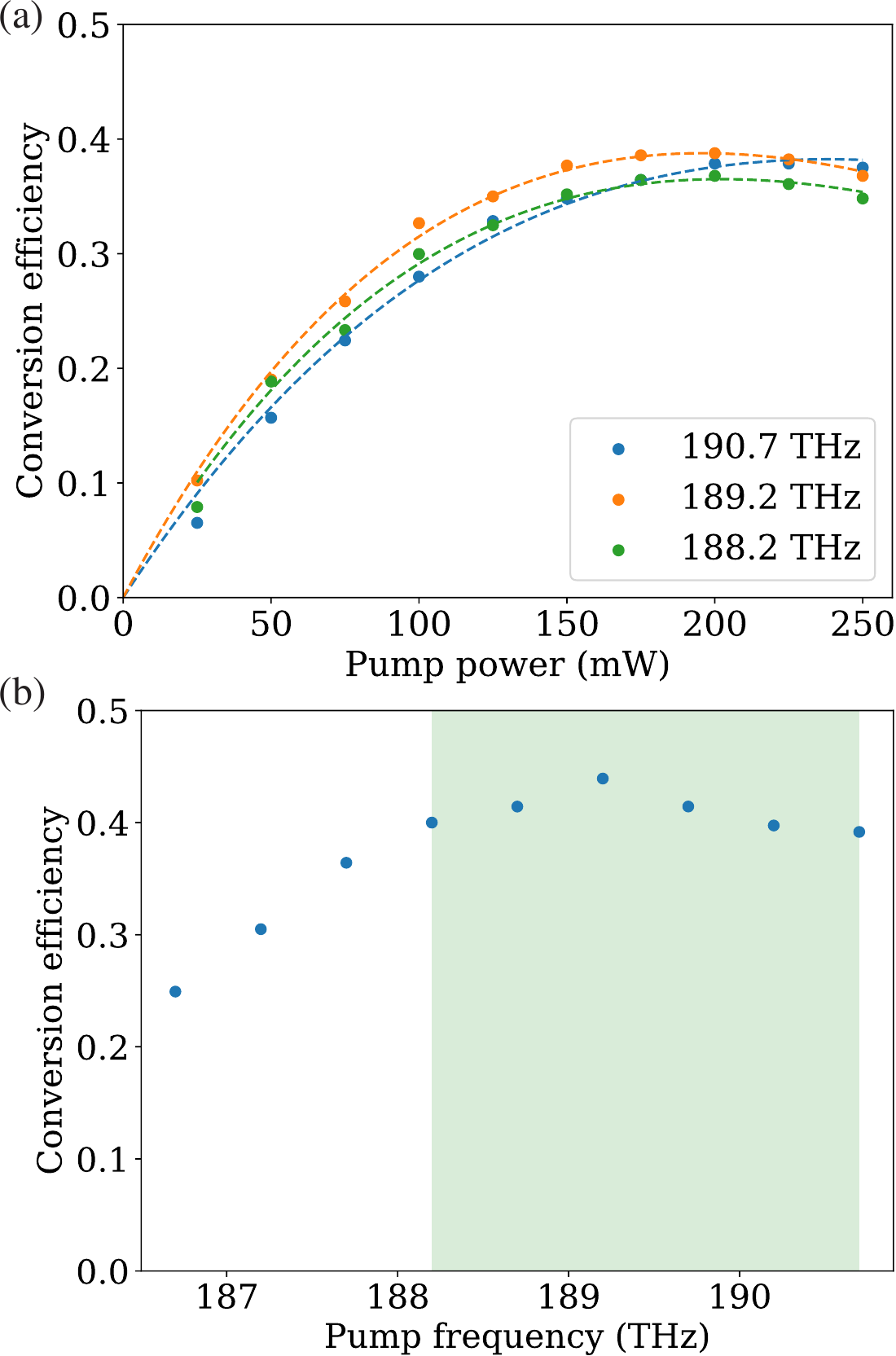}
  \caption{(a) Conversion efficiencies for pump lights 
  at the frequencies of 190.7, 190.2, and \SI{189.7}{THz}. 
  (b) The dependency of the maximum conversion efficiencies on the frequencies of the pump light.
    }
  \label{fig:conv}
\end{figure}
Before demonstrating the CS-QFC, 
we first investigated the conversion efficiencies of the device 
using continuous-wave laser light at \SI{780}{nm}. 
To measure the pump power dependence of the conversion efficiency, 
we directly connected one of the tunable pump laser with the EDFA. 
In the experiment, we used pump light at frequencies of 190.7, 189.2, and \SI{188.2}{THz}. 
The experimental results are shown in Fig.~\ref{fig:conv}(a). 
The best fit of the experimental data to a function of $A\sin^2(\sqrt{BP})$ gives 
$(A,B)=(0.38, \SI{0.010}{mW^{-1}})$, $(0.39, \SI{0.013}{mW^{-1}})$ and $(0.37, \SI{0.012}{mW^{-1}})$
for the pump frequencies of 190.7, 189.2 and \SI{188.2}{THz}, respectively. 
From the results, the pump power dependencies of the conversion efficiencies are similar across various pump frequencies. 
While the values of $B$ are not exactly equal, 
the passive adjustment of appropriate electric currents applied to the shutters of the pump channels followed by the EDFA at a constant gain 
can achieve the frequency conversion process 
with the maximum conversion efficiencies of $\eta\u{max}$ in each channel as we explained. 
By a similar logic, the conversion efficiency can be equalized across all channels.
Consequently, we conclude that the frequency conversion system 
is applicable to the large-scale frequency multiplexing. 

Next, we measured the dependence of the maximum conversion efficiencies on the frequencies of the pump light. 
We increased the pump frequencies in \SI{0.5}{THz} steps over a configurable range. 
In this experiment, the temperature of the PPLN-WG is optimized for the conversion using the pump light at \SI{189.2}{THz}.
The maximum conversion efficiency for each setting of the pump frequency was measured by fine-tuning the pump power in the vicinity of \SI{200}{mW} from the result in Fig.~\ref{fig:conv}(a). 
The experimental result of the dependence of the maximum conversion efficiencies is shown in Fig.~\ref{fig:conv}(b). 
We can see that in the range of the pump frequency 
\SI{188.2}-\SI{190.7}{THz} (shaded area), which corresponds to the pump wavelength range \SI{1592.9}-\SI{1572.1}{nm}, 
maximum conversion efficiencies of about \SI{40}{\%} or even higher are obtained. 
If we consider the above range of \SI{2.5}{THz} range as the acceptable range to be used as selectable channels for CS-QFC, 
the number of channels for DWDM is 100 with a channel spacing of \SI{25}{GHz}.
The wideband acceptable region of CS-QFC is comparable to 
the region of the signal and idler photon pairs generated by the SPDC process, which has been demonstrated in Ref.~\cite{Zhang2021}. 

\begin{figure}
  \centering
  \includegraphics[width=7cm]{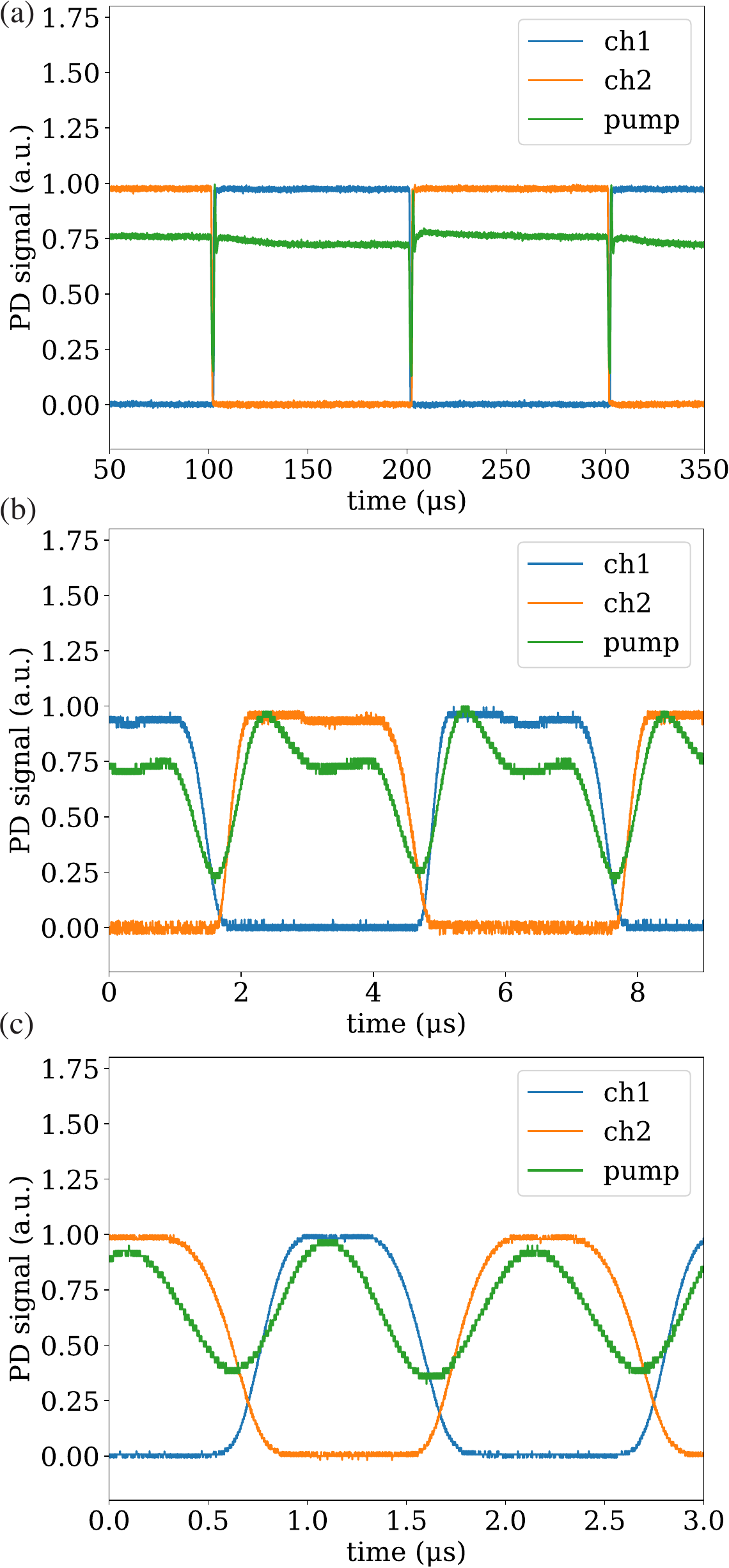}
  \caption{Observed signals of the converted light (blue: 194.5 THz, orange: 194.475 THz) and the pump light (green) observed during switching between the two pump lights at rates of (a) \SI{100}{\mu s}, (b) \SI{3}{\mu s}, and (c) \SI{1}{\mu s}.}
  \label{fig:switch}
\end{figure}
We characterized the switching property of the frequency conversion. 
We used two pump lasers at the frequencies of 189.731 and \SI{189.756}{THz}
with the detuning of \SI{25}{GHz}, which convert the signal light into 
the output frequency channel of 194.5 and \SI{194.475}{THz}, respectively. 
The corresponding converted wavelengths are 1541.35 and \SI{1541.55}{nm}. 
In the experiment, we switched the two pump lights 
every \SI{100}{\mu s}, \SI{3}{\mu s} or \SI{1}{\mu s}
and measured the converted light separated by the DeMux followed by the PDs and the digital sampling oscilloscope. 
The experimental result is shown in Fig.~\ref{fig:switch}. 
As a reference, we monitored the pump power reflected by the HPF, as represented by the green curves in Fig.~\ref{fig:switch}. 
For the case of the switching interval of \SI{100}{\mu s}, 
frequency-converted lights at the two channels 
are switched much faster than the switching interval. 
In contrast, for the cases of switching interval of \SI{3}{\mu s} and \SI{1}{\mu s}, 
we see the switching speed is comparable with the interval time. 
By determining the rise/fall time 
by the time it takes for the pulse to go from/to 10 to/from \SI{90}{\%}, 
the estimated rise and fall times are \SI{0.5}{\mu s}. 
Consequently, 
a switching interval slower than \SI{0.5}{\mu s} is required 
for the ideal operation of this device. 
It is important to note that the observed pump power shows a transient response 
when the pump light is turned on and off. 
We guess the reason might be that the light power stored in the EDFA is released at once. 
Despite the behavior of the pump light during switching, 
the converted light is insensitive to the transient response of the pump light. 
This is because the conversion efficiency determined by 
$A\sin(\sqrt{BP})$ is insensitive to changes in pump power 
around the maximum conversion efficiency. 

\subsection{CS-QFC using a single photon input}
\begin{figure}
  \centering
  \includegraphics[width=8.5cm]{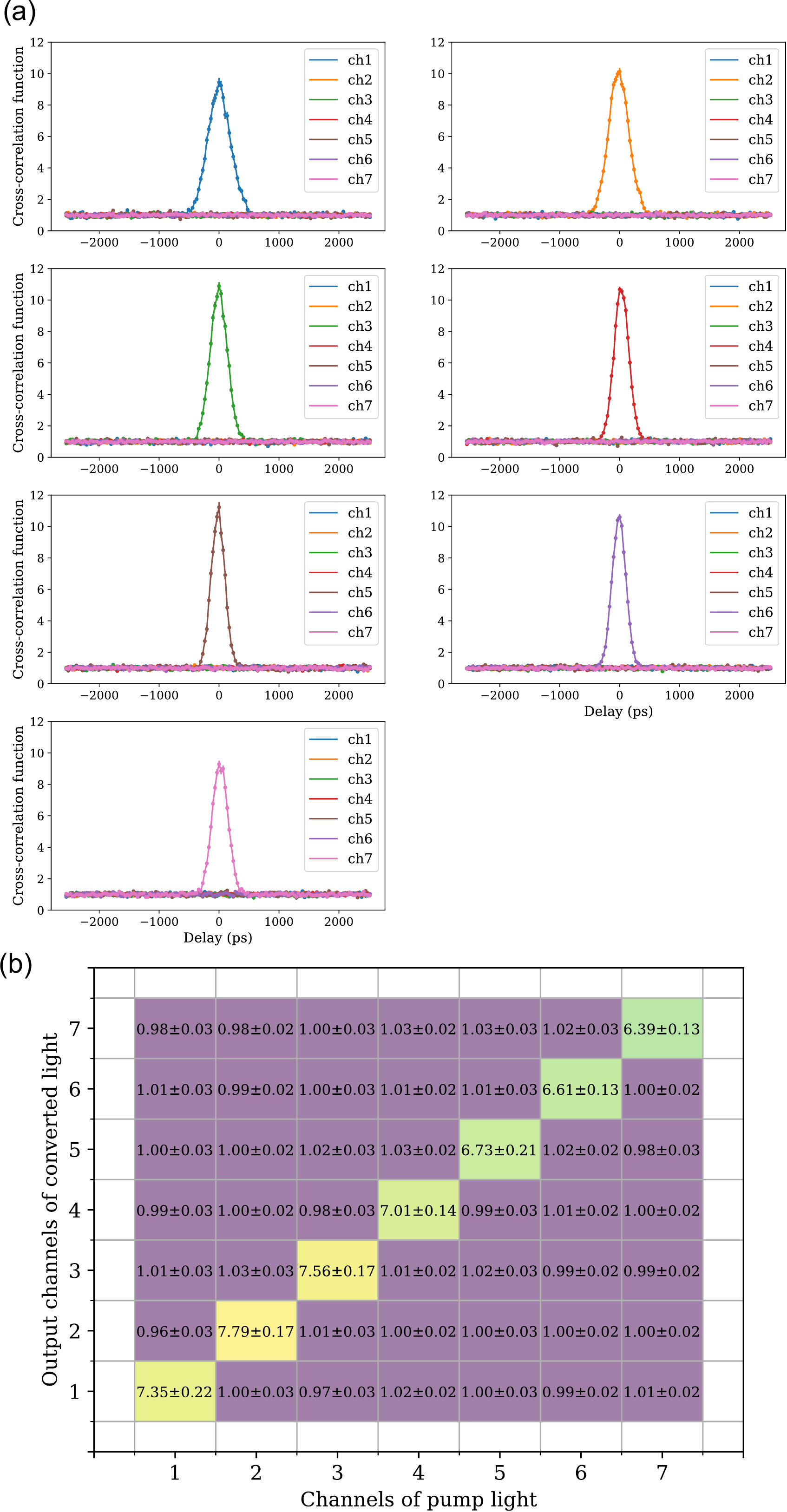}
  \caption{(a) The observed coincidence histograms of the time difference between the heralding photon and the converted photons. The coincidence counts are normalized as cross-correlation functions. (b) The estimated cross-correlation with the time window of \SI{476}{ps} in every output frequency channel for all pump frequencies.}
  \label{fig:coinc}
  \end{figure}

We performed the demonstration of CS-QFC 
using 7-ch pump lights 
at frequencies from 189.383 to \SI{189.533}{THz} with the step of \SI{25}{GHz}. 
The prepared SPDC photon pairs at 780 and \SI{1541}{nm} before QFC have the cross-correlation function of $18.22\pm0.04$. 
The \SI{780}{nm} heralded single photon is converted to the frequencies from 194.85 to \SI{194.7}{THz}, that are from channel 1 to channel 7 of the DeMux. 
To reduce the Raman scattering noises, we added the two bandpass filters with the bandwidths of \SI{12}{nm} and \SI{1.7}{nm} before the 1x16 DeMux. 
The experimental results of 
the observed cross-correlation functions between the heralding and converted photons are shown in Fig.~\ref{fig:coinc}(a). 
The accumulation time was \SI{180}{sec} and the width of the time bin was \SI{34}{ps}.
By using these histograms of Fig.~\ref{fig:coinc}(a) and the time window of \SI{476}{ps}, the estimated cross-correlations in each frequency channel 
for all pump frequencies are summarized in Fig.~\ref{fig:coinc}(b). 
As expected, a high cross-correlation function was observed 
in the output channel corresponding to the pump frequency 
due to the existence of the photons converted by the signal photons correlated with the heralding photon. 
This shows that the converted photons were observed only in the desired output channels 
without serious crosstalk on the other channels. 
We thus conclude the CS-QFC was successfully demonstrated. 
  
\section{Discussion}
In the experiment, we showed that the bandwidth of the CS-QFC is $\sim$\SI{2.5}{THz}, 
which corresponds to 100 switchable channels with a channel spacing of \SI{25}{GHz}. 
The bandwidth is determined by the acceptance bandwidth of the PPLN-WG. 
A shorter PPLN waveguide allows a larger number of switchable frequency channels, 
at the expense of a higher pump power required for the maximum conversion efficiency. 
While we have demonstrated the CS-QFC with only 7 output channels between \SI{175}{GHz}, 
at least 100-ch pump lasers, including the fast switches, are needed to effectively utilize the \SI{2.5}{THz} bandwidth. 
One way to achieve this with a low cost of the laser source preparation is 
to use of the wavelength-tunable laser as a pump light. 
According to the state of the art of ultrafast tunable laser~\cite{Snigirev2023}, 
electro-optic laser frequency tuning is possible at a rate of \SI{12}{PHz/s} while maintaining the narrow linewidth of $\sim$ kHz. 
Using this laser, we can take \SI{2}{\micro s} to tune the \SI{25}{GHz} frequency. 
For faster switching, multiple tunable lasers 
combined with a fast $N$-to-1 optical switch can be used. 
A commercially available $N$-to-1 switch~\cite{Nashimoto2005} 
has a switching speed of \SI{10}{ns}. 
It is sufficient to improve the switching speed of the ultrafast tunable laser described above.
Another way to reduce the cost of the light source is the use of an optical frequency comb split by DeMux with a spacing of \SI{25}{GHz} instead of 100-ch pump lasers.

As discussed above, there are several candidates for the pump light of CS-QFC. 
To apply the CS-QFC to quantum communication including quantum matter systems, 
not only the switching speed of the pump light but also the other properties such as 
the linewidth and stability are important to achieve 
high rate and high fidelity quantum communication. 
For the communication rate, 
the channel switching speed $\tau_s$ must be much faster than 
the temporal width $\tau_c$ of the photons is required 
in order not to decrease the rate. 
For the fidelity, because the indistinguishability of photons arrived from different nodes is crucial, 
the pump light with the frequency fluctuations degrades the visibility of the interference. 
Therefore, sufficient stability of the pump frequency is required. 
For example, a 780 nm photon entangled with a Rb atom is generated using D2 line transition with a linewidth of \SI{6}{MHz} and a temporal width of $\tau_c=\SI{26}{ns}$\cite{steck2001rubidium,Hofmann2012}.
In this case, the CS-QFC with a switching speed of $\tau_s \sim\SI{10}{ns}$ will not significantly affect the communication rate. 
For the fidelity, 
a sufficiently narrow linewidth and frequency stability much less than \SI{6}{MHz} is required for the pump light to keep the initial linewidth of the photon. 
This suggests multiple stable lasers or a frequency comb may 
be more promising than a conventional tunable laser for the pump light.
The above discussion is just an example of the quantum communication using Rb atoms. 
In practice, it is crucial to use a suitable pump light that meets the required switching speed and stability, tailored to the specific parameters of other quantum systems, multiplexing methods, and quantum communication protocols to be implemented.

\section{Conclusion}
We have demonstrated the CS-QFC, which converts a frequency 
of a signal photon to 
another frequency 
actively switchable
at each round of the conversion process. 
The switchable bandwidth of the CS-QFC is \SI{2.5}{THz}, 
achieving 100 frequency channels with a channel spacing of \SI{25}{GHz}. 
Using a heralded single photon generated by the SPDC, 
we successfully demonstrated the CS-QFC with the 7 output channels 
switchable by using one of the seven different frequencies of the pump light. 
Namely, the converted photons were observed in the desired output channel 
expected from the pump frequency, while preserving the non-classical photon statistics 
without the crosstalk to the other different channels. 
We have discussed the possibility of applying the CS-QFC to the conversion of 
photons emitted from the Rb atom 
by considering the candidates for the configuration of the pump light preparation 
using commercially available devices and state-of-the-art technologies.

The demonstrated CS-QFC provides switching functionality to the quantum interface, which selectively converts to the wavelengths in the WDM channels from a single frequency around the visible range of photons emitted from quantum matter systems. 
Combining CS-QFC and frequency converters on WDM channels~\cite{Joshi2018,Lee2004, Fisher2021,Wang2021,Yokota2022} allows for more flexible routing of the photons. This capability is valuable for realizing entanglement distribution with switching network designs~\cite{Drost2016,Koyama2024} and a DWDM-compatible quantum internet.

\section*{Acknowledgments}
This work was supported by Moonshot R \& D, JST JPMJMS2066, JST JPMJMS226C;  R \& D of ICT Priority Technology Project JPMI00316; FOREST Program, JST JPMJFR222V.
T.Y. and R.I. acknowledge the members of the Quantum Internet Task Force for the comprehensive and interdisciplinary discussions on the quantum internet.

\end{document}